\documentstyle[epsf]{aa}
\def\kms{\,km\thinspace s$^{-1}$ }  
\def\kmsk{\,km\thinspace s$^{-1}$ \thinspace kpc$^{-1}$}  
\begin{document}

\thesaurus{03(11.04.2;   
              11.05.1;   
              11.09.1;   
              11.09.2;   
              11.11.1;   
              11.19.6)}  

   \title{IC3328: a ``dwarf elliptical galaxy'' with spiral structure}

\author{Helmut Jerjen\,\inst{1,}
\thanks{Based on observations collected at the European Southern Observatory (ESO 63.O-0055)}
\and Agris Kalnajs\,\inst{1}
\and Bruno Binggeli\,\inst{2}} 

\offprints{H.~Jerjen, e-mail: jerjen@mso.anu.edu.au}   

\institute{Research School of Astronomy and Astrophysics, 
Australian National University, Private Bag, Weston Creek PO, ACT\,2611, Canberra, Australia
\and 
Astronomisches Institut, Universit\"at Basel, Venusstrasse 7,
CH-4102 Binningen, Switzerland}

\date{}

\maketitle
\markboth{Jerjen et al.: Dwarf elliptical galaxy with spiral structure}
{Jerjen et al.: Dwarf elliptical galaxy with spiral structure}  

\begin{abstract}
We present the 2-D photometric decomposition of the Virgo galaxy
\object{IC3328}. The analysis of the global light distribution of this morphologically 
classified nucleated dwarf elliptical galaxy (dE1,N) reveals a tightly 
wound, bi-symmetric spiral structure with a diameter of 4.5\,kpc, precisely 
centered on the nucleus of the dwarf. The amplitude of the spiral is
only three percent of the dwarf's surface brightness making it the faintest and 
smallest spiral ever found in a galaxy. In terms of pitch angle and arm winding 
the spiral is similar to the intermediate-type galaxy M51, but it lacks the dust 
and prominent \ion{H}{ii} regions which signal the presence of gas. The visual
evidence of a spiral pattern in an early-type dwarf galaxy reopens the question 
on whether these dwarfs are genuine rotationally supported or anisotropic stellar systems. 
In the case of IC3328, we argue for a nearly face-on disk (dS0) galaxy
with an estimated maximum rotation velocity of $v_{\mathrm{c,max}}\approx 55$\kms.
The faintness of the spiral and the small motions within it, suggests
that we could be seeing swing-amplified noise. The other possibility
is a tidal origin, caused by the near passage of a small companion.  

\end{abstract}

\section{Introduction}

\begin{figure*}
\centering\leavevmode
\epsfxsize=16cm
\epsfbox[43 435 536 683]{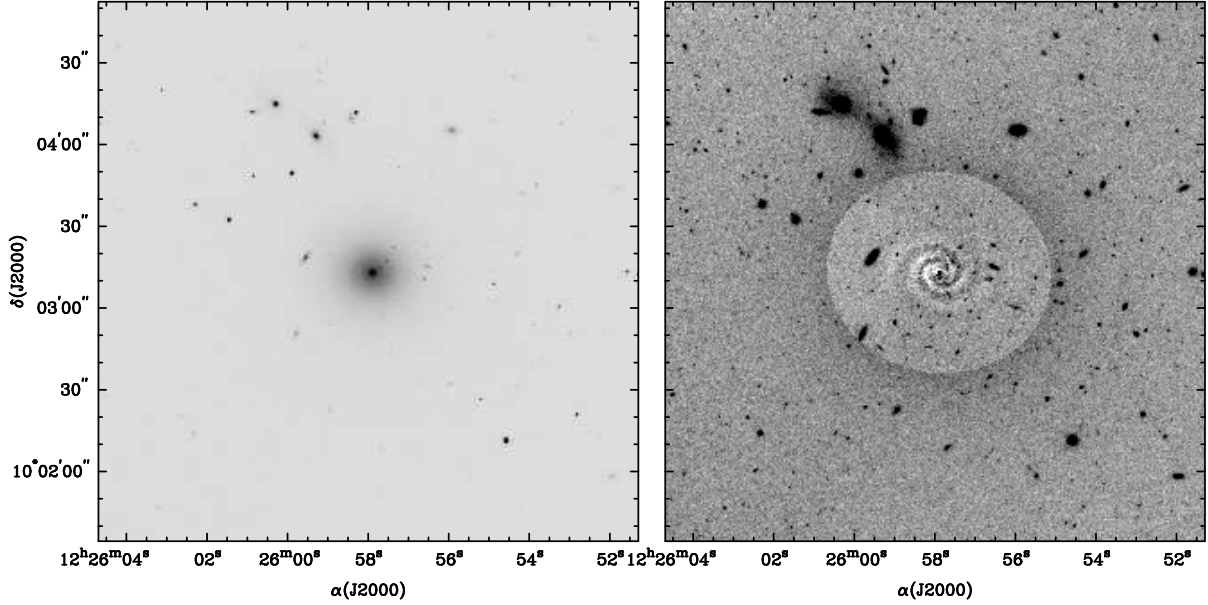}
\caption[jerjen.fig1.ps]{The deep R-band CCD image of IC3328 (left panel) 
illustrates the overall morphology of this as dE1,N classified galaxy: a 
smooth radially decreasing light distribution with a centrally located 
nucleus. After the subtraction of the axis symmetric component, the 
residual image (right panel) reveals a prominent 2-armed spiral structure 
with a possible central bar. 
\label{fig1}}
\end{figure*}

In this paper we report on the serendipitous discovery of a spiral structure
in the Virgo cluster dwarf elliptical IC3328. The presence of spiral structure
provides compelling evidence for the disk nature of that particular dwarf galaxy.
The observations are described and the light distribution is analysed in Sect.2. 
In Sect.3 we estimate the kinematical properties of the dwarf 
galaxy from the observed light distribution, assuming a likely 
value for the mass-to-light ratio and distance. We speculate 
on the origin of the spiral pattern in Sect.4. The concluding section 
deals with the ramifications for the dwarf elliptical taxonomy arising from IC3328.

\section{Photometric Evidence for a Spiral Structure}

IC3328 is known as an early-type dwarf galaxy in the 
Virgo cluster, morphologically classified as dE1,N (Virgo 
Cluster Catalog 856, Binggeli et al.~1985). 
The redshift of $v_\odot=972\pm 32$ \kms made this galaxy 
a probable cluster member and a good candidate for a more
refined distance determination based on the Surface 
Brightness Fluctuations (SBF) method. Candidates for 
the SBF method should have smooth and symmetric light 
distributions, and show no obvious dust or star-forming regions.

The first step in applying the SBF method is to determine the
mean 2-D surface brightness distribution by suitable averaging 
and then subtracting it from the galaxy image, leaving just the 
fluctuating part which arises from the Poisson distribution of 
the stellar sources, ie. the fluctuations due to unresolved stars 
in the galaxy. However in the case of IC3328 the residuals were 
most surprising. The composite of three 400 sec  $R$-band images 
of IC3328 obtained at the VLT with the FORS1 multi-mode instrument 
in Service Mode on July 13, 1999 under excellent seeing conditions 
($0.6\arcsec$) is shown in the left panel of Fig.\ref{fig1}. The surface 
brightness ``fluctuations'' obtained after subtracting the mean is 
shown in the right panel.

Clearly ``fluctuations'' is not the right word to describe the
exceptionally regular spiral whose amplitude is so low that it 
remained invisible in the original exposure. The spiral structure is 
confined to the inner $30\arcsec$ of the galaxy which has an isophotal 
radius of $r_{R, 27}$=80\arcsec . It appears that we have discovered
the real nature of IC3328: a seemingly dust-free disk galaxy with
unusually faint spiral structure.

\begin{figure}
\centering\leavevmode
\epsfxsize=8.5cm
\epsfbox[26 155 536 715]{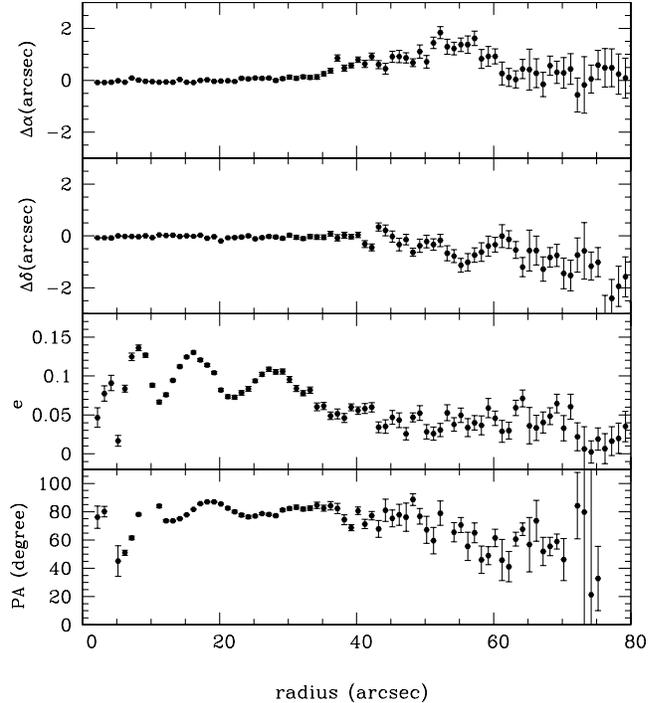}
\caption[jerjen.fig2.eps]{
The coordinate offset $\Delta \alpha$ and $\Delta \delta$ between 
the nucleus and the center of isophotal ellipses, ellipticity, and 
position angle (counterclockwise from north) shown as functions of 
radius. 
\label{fig2}}
\end{figure}

The initial analysis of the light distribution in IC3328 followed the
traditional route, based on standard IRAF procedures. The nuclear offsets
determined by fitting ellipses to the light distribution were perfectly
normal (top panels of Fig.\ref{fig2}). The first signs of unusual behaviour came
from the ellipticity and position angle variations of the ellipse fits. The 
coherent wiggles at smaller radii ($r<30\arcsec$) suggested interesting,
but low level structure, which could be seen as a spiral pattern in the
contour maps. 

\begin{figure}
\centering\leavevmode
\epsfxsize=8.5cm
\epsfbox[30 180 516 643]{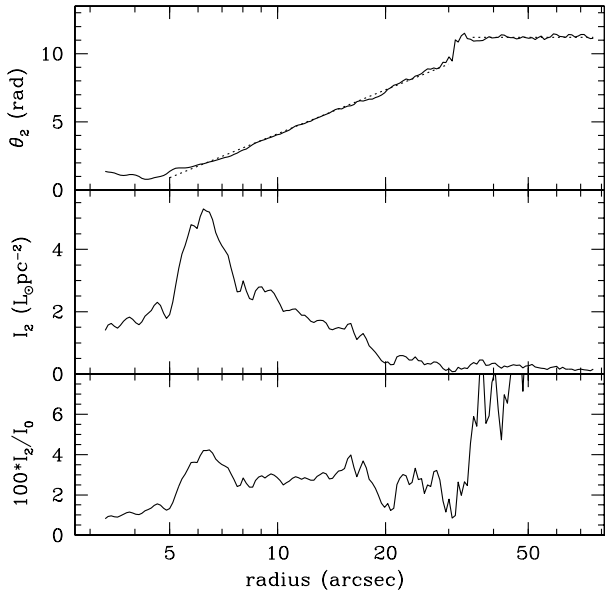}
\caption[jerjen.fig3.eps]{
Azimuthal phase (top panel) and amplitude (central panel) of the spiral 
as a function of radius. The logarithmic spiral approximation is shown
by the sloping dotted line. The fractional amplitude of the spiral is 
plotted in the bottom panel.
\label{fig3}}
\end{figure}

Contour maps, and in particular the smoothed contour maps reconstructed
from the ellipse fits are powerful tools for revealing faint structure,
but are not immediately suitable for quantitative analysis of the light
distribution. The presence of a smooth spiral indicates that we are
dealing with a disk or a disk embedded in a spheroidal mass distribution.
The most straight-forward analysis consists of guessing the orientation
of the disk and expanding the deprojected light distribution in a Fourier
series in the azimuthal angle $\theta$: 
\begin{eqnarray*}
&& I(r,\theta) =  I_0 (r) + I_1 (r) \cos [\theta - \theta_1 (r)]\\
&& \qquad \qquad +\; I_2 (r) \cos 2[\theta - \theta_2 (r)] + \cdots \ .
\end{eqnarray*}
The amplitude of the two-armed spiral, $I_2 (r)$, is quite sensitive
to changes in inclination, and less so to the position angle; a wrong
inclination will produce modulations in the amplitude. An inclination
of $25\fdg 2$ (which corresponds to the ellipticity $e=0.095$), and
the position angle of $82\fdg 5$ produced the smoothest two-armed
component. The amplitude and phase of this spiral is shown in
Fig.~\ref{fig3}. The fractional amplitude of $3-4\%$ is extremely low.

The one-armed, or $\cos \theta$ terms become important
close to the center ($ < 5\arcsec$).

In the interval $ 5\arcsec < r < 30\arcsec$ the phase $\theta_2 (r)$ is well
approximated by a straight line. Such a phase variation corresponds
to a two-armed logarithmic spiral inclined at $12 \fdg 1$ to a circle. 
The angular winding of the arms is $430^\circ$. In these
respects IC3328 resembles a Sb or Sbc galaxy very similar to M51
(Danver 1942; Kennicutt 1981), but without obvious gas, dust or 
bright \ion{H}{ii} regions.

\begin{figure}
\centering\leavevmode
\epsfxsize=8.5cm
\epsfbox[33 334 525 647]{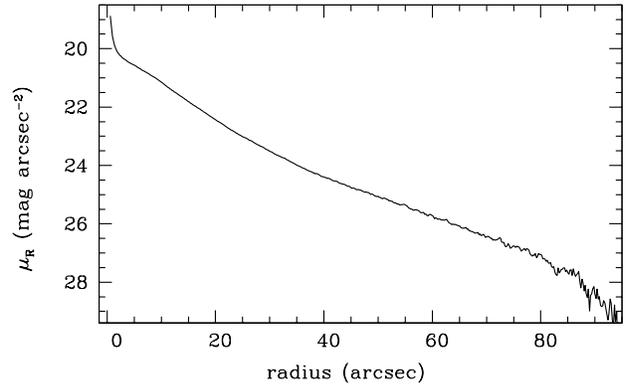}
\caption[jerjen.fig4.eps]{The $R$-band surface brightness 
profile of VCC0856 exhibits a distinct bi-linearity, characteristic 
for a type IIIb profile (see BC91). The transition occurs at 
$r\approx 30\arcsec$ precisely where the spiral fades out. 
\label{fig4}}
\end{figure}

The surface brightness profile derived from $I_0(r)$ is shown in
Fig.~\ref{fig4}. It can be approximated by two straight lines
(exponentials), with the cross-over occuring at $r \simeq 30\arcsec$,
which is also the place where the spiral pattern ends. The end
of the spiral can be seen in the flattening of the phase in
Fig.~\ref{fig3}. The $R$ surface brightness profile has the
same characteristics as the $B$ profile, which has been classified
as type IIIb by Binggeli \& Cameron (1991, hereafter BC91).

The total apparent magnitude, $R_T$, computed from $I_0(r)$ is $13.17$.
The half light radius $r_{\mathrm{eff}}=15\farcs 9$, and the mean surface
brightness within the effective radius $\langle \mu \rangle_{\mathrm{eff}}=21.17$\,mag
arcsec$^{-2}$. The best-fitting line (exponential) of the inner 
part has a central surface brightness of $\mu^{\mathrm{exp}}=19.95$\,mag
arcsec$^{-2}$ and a scale length $r^{\mathrm{exp}}=8 \farcs 7$. 

\section{Dynamics}

In order to estimate the dynamical time scales associated with the
spiral pattern we need to know the distance to and the velocities
within IC3328. 

There are two distance estimates for IC3328. The first comes from the
radial velocity which coincides with the mean velocity of the Virgo
cluster. It agrees well with the preliminary SBF distance of 15.5\,Mpc
(Jerjen et al. in preparation). At this distance $1\arcsec$ 
corresponds to $77.5$\,pc.

The only kinematic data available for IC3328 to date is a measurement
of the central velocity dispersion of $\sigma_{\mathrm{c}}=27$\kms (Peterson \&
Caldwell 1993). In the absence of a proper velocity field determination
we have to resort to the light distribution and estimates of the
mass-to-light ratio, $M/L$. 

\begin{figure}
\centering\leavevmode
\epsfxsize=8.5cm
\epsfbox[52 153 532 505]{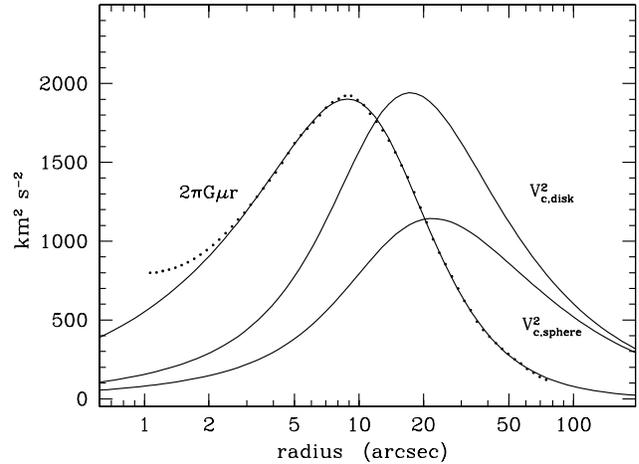}
\caption[jerjen.fig5.eps]{
A plot of $2\pi G \mu r$ and the two rotation curves, $V^2_{\mathrm{c,disk}}$
and $V^2_{\mathrm{c,sphere}}$, produced by it. The surface density $\mu (r)$
was obtained from $I_0(r)$ by assuming the $M/L=1$ (in solar units).
The $2\pi G \mu r$ curve is a Nuker profile fit to $2\pi G I_0(r) r$,
shown by the dots.
\label{fig5}}
\end{figure}

The square of the rotational velocity, $V_{\mathrm{c}}^2$, can be determined from
$2\pi G\mu r$, where $\mu(r)$ is the projected surface mass density. On 
a logarithmic scale these two quantities are related by a convolution 
(Kalnajs 1999). Fig.\ref{fig5} shows the relation between the two quantities 
in the case when $M/L=1$. The two $V_{\mathrm{c}}^2$ curves correspond to the limiting 
cases where the projected surface density comes from a flat or a spherical 
mass distribution. Fig.\ref{fig5} also makes it clear that the value of 
$M/L$ around the peak $2\pi G\mu r$ is what really matters.

Assuming $M/L=1$ gives a maximum disk rotation velocity, $V_{\mathrm{c,max}}=44$\kms.
A better estimate of $55$\kms comes from the average mass-to-light 
ratio for globular clusters $(M/L)_R=1.6$, based on the quantities
$(M/L)_V=2.5$ (Pryor \& Meylan 1993) and $(V-R)_0=0.47$ (Peterson 1993).
Such a value does not clash with $\sigma_{\mathrm{c}}=27$\kms.

The best option would be to measure the rotation curve. Then one could use
the above arguments to obtain the actual $M/L$ ratio of IC3328.

The estimated peak rotation velocity of $55$\kms occurs around $1.4$ kpc, which
means that the angular rotation rate there is $39$\kmsk, a value
comparable to the $25$\kmsk measured near the Sun. Thus IC3328 has
had ample time to settle into an equilibrium.

\section{Spiral Structure}

The presence of the spiral implies the presence of a disk. If what we
see is a nearly face-on disk then the small spiral amplitude is quite
unusual. We can estimate the mean displacements and velocities that
are needed to produce the observed density contrast.

A density wave is created by coherent oscillations of stars around 
their equilibrium orbits. Because the spiral is tightly wrapped, the
largest contribution comes from the radial displacements, $\delta r$,
whose phases are rapidly varying functions of the equilibrium position
$r$. If we write the phase as $\exp\{i[\alpha \ln(r) - 2\theta]\}$, then a
little bit of algebra shows that the change in surface density, $\delta \mu$
satisfies the relation
$$
|\frac{\delta \mu}{\mu}| \simeq |\alpha \frac{\delta r}{r}|   
$$
when $\alpha \gg 1$.
For our spiral $\alpha = \frac{2}{\sin 12\fdg 1} = 9.5$, which means that
when $\frac{\delta \mu}{\mu} \simeq 0.03$, $\frac{\delta r}{r} \simeq 0.003$,
a very small number indeed!

In a similar spirit we can estimate the velocities, $\delta V$, associated
with such displacements. The radial displacements will oscillate with
a frequency typically less than the natural radial frequency $\kappa$,
which in turn is roughly $\sqrt{2}$ times the mean angular rotation
rate $\Omega$. Hence $\frac{\delta V}{V_{\mathrm{c}}} \simeq 0.005$, or
$0.3$\kms.

These small numbers can be increased by reducing the disk light
contribution and attributing it to a spheroid. Even a ten-fold reduction
would produce fractional displacements of $3\%$, and velocities of
only $3$\kms. Such low velocities are unlikely to lead to shocks
in any neutral gas that may be present in the disk, and therefore
to any star formation that would signal the presence of gas.

\begin{figure}
\centering\leavevmode
\epsfxsize=8.5cm
\epsfbox[173 269 438 523]{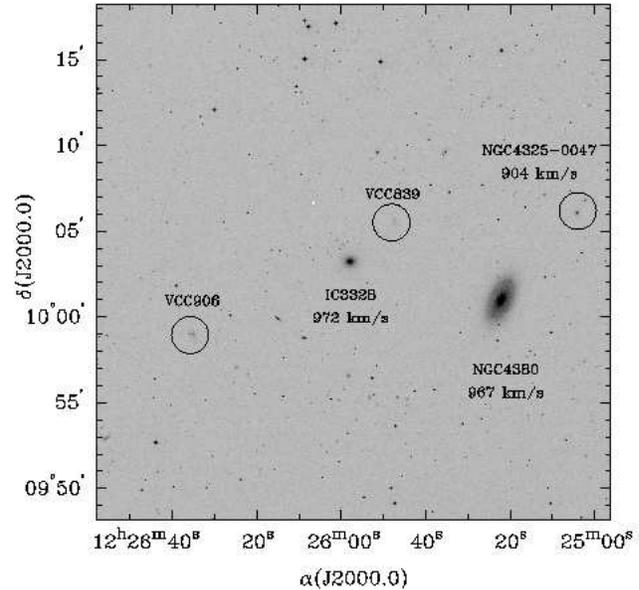}
\caption[jerjen.fig6.ps]{
A DSS-II image centered at IC3328 showing the galaxy distribution within a 15
arcmin squared field. The bright spiral galaxy NGC4380 is at a distance of 9.1 arcmin or 46
kpc to the West. Slightly further away is the faint galaxy NGC4325-0047. 
Furthermore, there are the two hardly visible, low surface brightness dwarf 
galaxies VCC839 (dE0:) and VCC906 (dE:) indicated by circles. Heliocentric 
velocities are given if available.
\label{fig6}}
\end{figure}

By reducing the disk light contribution we also reduce the importance
of the spiral's self-gravity and increase the likelihood of a tidal
origin. Fig.\ref{fig6} shows several possible perturbers. When the
self-gravity becomes negligible, one can rule out even a distant
passage of a big perturber like \object{NGC4380}, since the tidal distortions
will not propagate to the center. That leaves close passages by the
fainter objects, such as the two VCC dwarf galaxies, as possible cause
of the spiral seen in IC3328.

The other plausible scenario is that most of the light does come
from the disk and we are seeing swing amplified noise (Toomre \&
Kalnajs 1991). The gain of the swing amplifier is not large enough
to amplify the $\sqrt{N}$ stellar density fluctuations to the
observed $3\%$ level, but a small amount lumpy gas could provide
the necessary leading perturbations which then are amplified by a
factor of $10-30$ as the shear transforms them into a trailing
spirals (Toomre 1981). Huchtmeier \& Richter's (1986) upper limit 
of $\approx 6\cdot 10^7 M_{\sun}$ for the \ion{H}{i} content of 
IC3328 does not rule out this possibility.

\section{On the Taxonomy of Dwarf Elliptical Galaxies}
Dwarf galaxies come in two basic brands: dwarf ellipticals (dEs) and dwarf
irregulars (dIrrs). The common view is that dEs are spheroids and dIrrs are
disks. In the case of irregulars, the disk nature is clearly indicated by the
typical rotation pattern found in HI radio data (at least for dIrrs more
luminous than $M_B = -12$). Owing to the absence of gas and hence the lack 
of an easily accessible kinematic tracer, the situation is less clear for 
dEs. The basic evidence for their spheroidal nature is purely statistical: the 
flattening distribution of this dwarf type is very similar to the one
shown by normal ellipticals, which is distinctly different from disk galaxies
(Ryden \& Terndrup 1994; Binggeli \& Popescu 1995). Supporting evidence
for the spheroidal nature of dEs comes from spectroscopy. The measurements
so far clearly show a lack of rotation (Paltoglou \& Freeman 1987; Bender 
\& Nieto 1990; Bender at al.~1991).

However, there are strong indications suggesting that at least {\em part}\/ of 
the dwarf galaxies classified ``dE'' might be genuine disk galaxies:

(1) There is a general photometric kinship between dEs and low-luminosity
late-type galaxies (dIrrs and spirals of type Sd and Sm); e.g. both have 
roughly exponential surface brightness profiles (e.g. Lin \& Faber 1983;
Binggeli \& Cameron 1993). Coupled with the general
morphology-density relation of dwarf galaxies (Einasto et al.~1974; Binggeli
et al.~1987), this has fostered the idea that a significant fraction of the
dwarf ellipticals might be the dead ends of evolved irregulars and 
low-luminosity spirals (Lin \& Faber 1983; Kormendy 1985; Ferguson \& 
Sandage 1989; see Ferguson \& Binggeli 1994 for a broader discussion). 
In case of a sufficiently soft evolutionary mechanism such as ram pressure 
stripping, a dwarf galaxy could indeed have preserved its disk structure
while loosing its gas. 

(2) Sandage \& Binggeli (1984), in their morphological work on Virgo cluster
dwarfs, introduced the class of dwarf S0 galaxies. Objects classified dS0,
while being much rarer than dEs and confined to a bright magnitude range, are
characterized either by a hint of a two-component (S0-like) structure,
or some other peculiar feature like boxyness, bar-like feature, or extreme 
flattening (see also BC91). The flattening distribution of dS0s is typical 
for disk galaxies (Binggeli \& Popescu 1995). 

(3) More evidence for the structural diversity of dEs is provided
by Ryden et al. (1999). From an isophotal analysis of a sample of Virgo 
cluster dEs and dS0s these authors found the same range and frequency of
``boxy'' versus ``disky'' distortions from ellipticity as in normal 
ellipticals (e.g. Bender et al.~1989). Clearly, some of these early-type 
dwarf galaxies will be disk galaxies, or at least have a disk component
embedded.
 
The fourth and latest indication for a disk in a dE of course is the 
spiral in IC3328. Despite its classification as dE, this galaxy clearly
is a mis-classified dS0. We would like to emphasize that IC3328 does not 
resemble a ``dwarf spiral'', a new galaxy type proposed by Schombert 
et al.~(1995), as it lacks a bulge and shows no obvious signs of gas and dust.

It is very likely that more dS0s will be identified among the bright round
dEs in the future by means of a careful study of the 2-D light distribution.

\begin{acknowledgements}
The authors thank the anonymous ESO service observer at the VLT UT1 (ANTU) for 
providing excellent quality images and Ken Freeman for interesting discussions. 
The anonymous referee made helpful comments and suggestions. H.J. and B.B. are 
grateful to the {\em Swiss National Science Foundation} for financial support. 
\end{acknowledgements}

\end{document}